\documentclass[aps,preprint]{revtex4}%
\usepackage{amsfonts}
\usepackage{amsmath}
\usepackage{amssymb}
\usepackage{graphicx}%
\begin{document}
\preprint{CTP-SCU/2019006}
\title{Thermodynamics and Weak Cosmic Censorship Conjecture in Nonlinear
Electrodynamics Black Holes via Charged Particle Absorption}
\author{Peng Wang}
\email{pengw@scu.edu.cn}
\author{Houwen Wu}
\email{iverwu@scu.edu.cn}
\author{Haitang Yang}
\email{hyanga@scu.edu.cn}
\affiliation{Center for Theoretical Physics, College of Physical Science and Technology,
Sichuan University, Chengdu, 610064, PR China}

\begin{abstract}
We first obtain the $D$-dimensional asymptotically AdS charged black hole
solution in general nonlinear electrodynamics (NLED). We then use the
Hamilton-Jacobi method to describe the motion in curved spacetime of a scalar
particle and a fermion, which, as we show, satisfy the same Hamilton-Jacobi
equation. With absorbing a charged particle, the variation of the generic
charged NLED black hole is calculated in the normal and extended phase spaces.
In the normal phase space, where the cosmological constant and dimensionful
parameters in NLED are fixed, the first and second laws of thermodynamics are
satisfied. In the extended phase space, where the cosmological constant and
dimensionful parameters in NLED are treated as thermodynamic variables, the
first law of thermodynamics is also satisfied. However, the black hole entropy
can either increase or decrease depending on the changes in the dimensionful
parameters. Furthermore, we find that the weak cosmic censorship conjecture is
valid for the extremal and near-extremal black holes in the both phase spaces.

\end{abstract}
\keywords{}\maketitle
\tableofcontents

\bigskip



\section{Introduction}

Explaining the thermodynamic nature of black holes can have a deep impact upon
the understanding of quantum gravity. By making a black hole with an
ergosphere absorb a particle with negative energy, Penrose noted that energy
can be extracted from the black hole \cite{IN-Penrose:1969pc}. Later, an
irreducible quantity, namely the irreducible mass, was found to exist when a
black hole absorbs a particle
\cite{IN-Christodoulou:1970wf,IN-Bardeen:1970zz,IN-Christodoulou:1972kt}. Due
to the similarity between the irreducible mass and the entropy, Bekenstein
proposed that the area of a black hole (or the square of its irreducible mass)
corresponds to its entropy \cite{IN-Bekenstein:1972tm,IN-Bekenstein:1973ur}.
Moreover, the four laws of black hole mechanics were established in
\cite{IN-Bardeen:1973gs}, which are analogous to the laws of thermodynamics.
Without using any complete quantum theory of gravity, the semi-classical
approach allows for the computation of the thermodynamic quantities of black
holes. Indeed, with the advent of quantum field theory in curved spacetime,
Hawking found that black holes possess the temperature
\cite{IN-Hawking:1974sw}.

Investigating the thermodynamic properties of black holes has become quite a
mature subject over the years. In particular, the study of asymptotically
anti-deSitter black holes has become important since the discovery of the
AdS/CFT correspondence \cite{IN-Maldacena:1997re}. Hawking and Page showed
that a first-order phase transition occurs between the Schwarzschild AdS black
hole and the thermal AdS space \cite{IN-Hawking:1982dh}, which was later
understood as a confinement/deconfinement phase transition in the context of
the AdS/CFT correspondence \cite{IN-Witten:1998zw}. There has been much
interest in studying the thermodynamics and phase structure of AdS black holes
\cite{IN-Chamblin:1999tk,IN-Chamblin:1999hg,IN-Caldarelli:1999xj,IN-Cai:2001dz,IN-Kubiznak:2012wp}%
. Recently, the asymptotically AdS black holes have also been studied in the
context of the extended phase space thermodynamics, where the cosmological
constant is interpreted as a thermodynamic pressure
\cite{IN-Kastor:2009wy,IN-Dolan:2011xt}. The $P$-$V$ criticality study and
thermodynamics have been explored for various AdS black holes in the extended
phase space
\cite{IN-Wei:2012ui,IN-Gunasekaran:2012dq,IN-Cai:2013qga,IN-Xu:2014kwa,IN-Frassino:2014pha,IN-Dehghani:2014caa,IN-Hennigar:2015esa,IN-Wang:2018xdz}%
, where some intriguing phase behavior, e.g., reentrant phase transitions,
tricritical points, was found to be present. Therefore, studying the
thermodynamics and phase structure of black holes in both the normal and
extended phase spaces is important and interesting.

The weak cosmic censorship conjecture asserts that singularities are always
hidden by event horizons in any real physical process from regular initial
conditions and cannot be seen by the observers at the future null infinity
\cite{IN-Penrose:1969pc}. To test the validity of the weak cosmic censorship
conjecture, Wald attempted to overcharge or overspin an extremal Kerr-Newman
black hole by throwing a test particle into the black hole \cite{IN-Wald}.
However, for the particle with sufficient charge or angular momentum to
overcharge or overspin the black hole, the black hole can not capture it
because of the electromagnetic or centrifugal repulsion force. On the other
hand, it was found that near-extremal charged/rotating black holes could be
overcharged/overspun by absorbing a particle
\cite{IN-Hubeny:1998ga,IN-Jacobson:2009kt,IN-Saa:2011wq}. However, subsequent
researches showed that the weak cosmic censorship conjecture might be still
valid for these black holes with consideration of the backreaction and
self-force effects
\cite{IN-Hod:2008zza,IN-Barausse:2010ka,IN-Barausse:2011vx,IN-Zimmerman:2012zu,IN-Colleoni:2015afa,IN-Colleoni:2015ena}%
. Since there is still a lack of the general proof of the weak cosmic
censorship conjecture, its validity has been tested in various black holes
\cite{IN-Matsas:2007bj,IN-Richartz:2008xm,IN-Isoyama:2011ea,IN-Gao:2012ca,IN-Hod:2013vj,IN-Duztas:2013wua,IN-Siahaan:2015ljs,IN-Natario:2016bay,IN-Duztas:2016xfg,IN-Revelar:2017sem,IN-Sorce:2017dst,IN-Husain:2017cmj,IN-Crisford:2017gsb,IN-An:2017phb,IN-Ge:2017vun,IN-Yu:2018eqq,IN-Gwak:2018akg,IN-Gim:2018axz,IN-Chen:2018yah,IN-Zeng:2019jta,IN-Chen:2019nsr,IN-Gwak:2019asi,IN-Zeng:2019jrh,IN-Chen:2019pdj}%
. In particular, the thermodynamics and the weak cosmic censorship conjecture
have been considered for a $D$-dimensional Reissner-Nordstrom (RN)-AdS black
via the charged particle absorption in the extended phase space
\cite{IN-Gwak:2017kkt}. It showed that the first law of thermodynamics and the
weak cosmic censorship conjecture are satisfied while the second law of
thermodynamics is violated near extremality. It is noteworthy that the second
law of thermodynamics is always valid for a RN-AdS black hole in the normal
phase space.

If quantum corrections are considered, nonlinear terms are usually added to
the Maxwell Lagrangian, which is an effective model, namely the nonlinear
electrodynamics (NLED). In the Einstein-NLED theory, various NLED charged
black hole solutions were derived and discussed in a number of papers
\cite{IN-Soleng:1995kn,IN-AyonBeato:1998ub,IN-Maeda:2008ha,IN-Fan:2016hvf,IN-Fan:2016rih,IN-Hendi:2017mgb,IN-Tao:2017fsy,IN-Guo:2017bru,IN-Mu:2017usw,IN-Kuang:2018goo}
(for a brief review see \cite{IN-Bronnikov:2017xrt}). A globally regular NLED
black hole solution with nonzero electric charge can be constructed, which
requires that the NLED Lagrangian is strongly non-Maxwell in the weak field
limit \cite{IN-Bronnikov:2000vy,IN-Bronnikov:2000yz}. Specifically, the
thermodynamics of NLED black holes has been considered in the literature,
e.g., power Maxwell invariant black holes \cite{IN-Hendi:2012um,IN-Mo:2016jqd}%
, non-linear magnetic-charged dS black holes \cite{IN-Nam:2018tpf},
Born-Infeld AdS black holes
\cite{IN-Fernando:2003tz,IN-Fernando:2006gh,IN-Banerjee:2010da,IN-Banerjee:2011cz,IN-Lala:2011np,IN-Banerjee:2012zm,IN-Zou:2013owa,IN-Azreg-Ainou:2014twa,IN-Hendi:2015hoa,IN-Zangeneh:2016fhy,IN-Zeng:2016sei,IN-Li:2016nll,IN-Hendi:2017oka,IN-Dehyadegari:2017hvd,IN-Momennia:2017hsc}%
, Born-Infeld black holes in a cavity \cite{IN-Wang:2019kxp}.

In this paper, we extend the analysis of \cite{IN-Gwak:2017kkt} (a RN-AdS
black hole) to study the thermodynamics and the weak cosmic censorship
conjecture for a $D$-dimensional AdS charged black hole in a general
Einstein-NLED theory. Our results are NLED model independent and show that the
results of \cite{IN-Gwak:2017kkt} are quite robust against the corrections
from NLED. The rest of this paper is organized as follows. In section
\ref{Sec:HJE}, we derive the the Hamilton-Jacobi equation for a particle in
curved spacetime and discuss its motion around the black hole horizon. In
section \ref{Sec:NLEDBH}, we obtain general form of the $D$-dimensional NLED
charged AdS black hole solution. In section \ref{Sec:TCPA}, the thermodynamics
of the black hole are discussed in the normal and extended phase space. We
test the weak cosmic censorship conjecture in section \ref{Sec:WCCC}. We
summarize our results in section \ref{Sec:Con}. For simplicity, we set
$G=\hbar=c=k_{B}=1$ in this paper.

\section{Hamilton-Jacobi Equation}

\label{Sec:HJE}

In \cite{HJE-Benrong:2014woa}, we have already derived the Hamilton-Jacobi
equations for a scalar particle and a fermion in a curved spacetime background
under an electric potential $A_{\mu}$ and showed that these Hamilton-Jacobi
equations have the same form. Here, we first briefly review the derivation of
the Hamilton-Jacobi equation. After the Hamilton-Jacobi equation is obtained,
the motion of the particle near the horizon of a charged black hole is
discussed. In this section, we temporarily restore $\hbar$ to keep the track
of the leading term of the WKB expansion.

\subsection{Scalar Particle}

In a semiclassical approximation, the Hamilton-Jacobi equation for a scalar
particle is the lowest order of the WKB expansion of the corresponding
Klein-Gordon equation. In curved spacetime, for a scalar particle of the mass
$m$ and the charge $q$, the Klein-Gordon equation is%

\begin{equation}
D^{\mu}D_{\mu}\phi+\frac{m^{2}}{\hbar^{2}}\phi=0\text{,} \label{eq:KG}%
\end{equation}
where $D_{\mu}=\partial^{\mu}-iqA^{\mu}/\hbar$ is the covariant derivative,
and $A_{\mu\text{ }}$is the electromagnetic potential. Assuming the WKB ansatz
for $\phi$:%
\begin{equation}
\phi=\exp\left(  \frac{iI}{\hbar}\right)  ,
\end{equation}
one can expand eqn. $\left(  \ref{eq:KG}\right)  $ in powers of $\hbar$ and
finds that the lowest order term is
\begin{equation}
\left(  \partial^{\mu}I-qA^{\mu}\right)  \left(  \partial_{\mu}I-qA_{\mu
}\right)  =m^{2}. \label{eq:Hamilton-JacobiS}%
\end{equation}
which is the Hamilton-Jacobi equation.

\subsection{Fermion}

In curved spacetime, the Dirac equation for a spin-$1/2$ fermion of the mass
$m$ and the charge $q$ takes on the form as%
\begin{equation}
i\gamma_{\mu}\left(  \partial^{\mu}+\Omega^{\mu}-\frac{iq}{\hbar}A^{\mu
}\right)  \psi-\frac{m}{\hbar}\psi=0, \label{eq:Dirac}%
\end{equation}
where $\Omega_{\mu}\equiv\frac{i}{2}\omega_{\mu}^{\text{ }ab}\Sigma_{ab}$,
$\Sigma_{ab}$ is the Lorentz spinor generator, $\omega_{\mu}^{\text{ }ab}$ is
the spin connection and $\left\{  \gamma_{\mu},\gamma_{\nu}\right\}
=2g_{\mu\nu}$. The Greek indices are raised and lowered by the curved metric
$g_{\mu\nu}$, while the Latin indices are governed by the flat metric
$\eta_{ab}$. To obtain the Hamilton-Jacobi equation for the fermion, the WKB
ansatz for $\psi$ is assumed as
\begin{equation}
\psi=\exp\left(  \frac{iI}{\hbar}\right)  v, \label{eq:fermionansatz}%
\end{equation}
where $v$ is a slowly varying spinor amplitude. Substituting eqn. $\left(
\ref{eq:fermionansatz}\right)  $ into eqn. $\left(  \ref{eq:Dirac}\right)  $,
we find that the lowest order term of $\hbar$ is%
\begin{equation}
\gamma_{\mu}\left(  \partial^{\mu}I-qA^{\mu}\right)  v=-mv,
\label{eq:Hamilton-JacobiF}%
\end{equation}
which is the Hamilton-Jacobi equation for the fermion. Multiplying both sides
of eqn. $\left(  \ref{eq:Hamilton-JacobiF}\right)  $ from the left by
$\gamma_{\nu}\left(  \partial^{\nu}I+eA^{v}\right)  \,$and then using eqn.
$\left(  \ref{eq:Hamilton-JacobiF}\right)  $ to simplify the RHS, one obtains%
\begin{equation}
\gamma_{\nu}\left(  \partial^{\nu}I-qA^{\nu}\right)  \gamma_{\mu}\left(
\partial^{\mu}I-qA^{\mu}\right)  v=m^{2}v.
\end{equation}
Using $\left\{  \gamma_{\mu},\gamma_{\nu}\right\}  =2g_{\mu\nu}$, we have%
\begin{equation}
\left[  \left(  \partial^{\mu}I-qA^{\mu}\right)  \left(  \partial_{\mu
}I-qA_{\mu}\right)  -m^{2}\right]  v=0.
\end{equation}
Since $v$ is nonzero, the Hamilton-Jacobi equation reduces to%
\begin{equation}
\left(  \partial^{\mu}I-qA^{\mu}\right)  \left(  \partial_{\mu}I-qA_{\mu
}\right)  =m^{2},
\end{equation}
which is the same as the Hamilton-Jacobi equation for a scalar.

\subsection{Motion near Black Hole Horizon}

To include a broader class, we consider a $D$-dimensional static black hole
with the line element
\begin{equation}
ds^{2}=-f\left(  r\right)  dt^{2}+\frac{1}{f\left(  r\right)  }dr^{2}+C\left(
r^{2}\right)  h_{ab}\left(  x\right)  dx^{a}dx^{b}, \label{eq:BHmetric}%
\end{equation}
where $a$, $b=1,\cdots D-2$. We also assume that there is a presence of
electromagnetic potential $A_{\mu}$ with the following ansatz%
\begin{equation}
A_{\mu}=A_{t}\left(  r\right)  \delta_{\mu t},
\end{equation}
which means that the black hole is electrically charged. Suppose that the
outermost horizon of the black hole is at $r=r_{+}$, which gives that
$f\left(  r_{+}\right)  =0$ and $f\left(  r\right)  >0$ for $r>r_{+}$.

In the metric $\left(  \ref{eq:BHmetric}\right)  $, the Hamilton-Jacobi
equation $\left(  \ref{eq:Hamilton-JacobiS}\right)  $ reduces to%
\begin{equation}
-\frac{\left(  \partial_{t}I-qA_{t}\right)  ^{2}}{f\left(  r\right)
}+f\left(  r\right)  \left(  \partial_{r}I\right)  ^{2}+\frac{h^{ab}\left(
x\right)  \partial_{a}I\partial_{b}I}{C\left(  r^{2}\right)  }=m^{2}\text{,}%
\end{equation}
where $h^{ab}h_{bc}=\delta_{c}^{a}$. Since $I$ is the classical action,
$E=-\partial_{t}I$ is the conserved energy of the particle. To solve the
Hamilton-Jacobi equation, we can employ the following ansatz%
\begin{equation}
I=-Et+W\left(  r\right)  +\Theta\left(  x\right)  .
\end{equation}
The method of separation of variables gives the differential equation for
$\Theta\left(  x\right)  $%
\begin{equation}
h^{ab}\left(  x\right)  \partial_{a}\Theta\left(  x\right)  \partial_{b}%
\Theta\left(  x\right)  =\lambda, \label{eq:HJh}%
\end{equation}
where is $\lambda$ is a constant. On the other hand, the differential equation
for $W\left(  r\right)  $ can lead to
\begin{equation}
E=-qA_{t}\left(  r\right)  +\sqrt{f\left(  r\right)  \left[  m^{2}%
+\frac{\lambda}{C\left(  r^{2}\right)  }\right]  +\left[  P^{r}\left(
r\right)  \right]  ^{2}}, \label{eq:energyofparticle}%
\end{equation}
where $P^{r}\left(  r\right)  \equiv f\left(  r\right)  \partial_{r}W\left(
r\right)  $ is the radial momentum of the particle, and we choose the positive
sign in front of the square root since the energy of the particle is required
to be a positive value \cite{IN-Christodoulou:1970wf,IN-Christodoulou:1972kt}.
As noted in \cite{HJE-Tao:2017mpe}, $P^{r}\left(  r\right)  $ is finite and
nonzero at $r=r_{+}$, which accounts for the Hawing radiation modeled as a
tunneling process. So very close to the outer horizon at $r=r_{+}$, eqn.
$\left(  \ref{eq:energyofparticle}\right)  $ becomes%
\begin{equation}
E=q\Phi+\left\vert P^{r}\left(  r_{+}\right)  \right\vert ,
\label{eq:Ehorizon}%
\end{equation}
where $\Phi\equiv-A_{t}\left(  r_{+}\right)  $ is the electric potential of
the black hole. Eqn. $\left(  \ref{eq:energyofparticle}\right)  $ relates the
energy of the particle to the momentum and the potential energy near $r=r_{+}$.

\section{$D$-dimensional NLED Black Hole Solution}

\label{Sec:NLEDBH}

For general NLED theories, the four-dimensional static and spherically
symmetric black hole solution with an electric field has already been given in
\cite{NLEDBH-Pellicer:1969cf,NLEDBH-Pleb:1969}. Here, we extend the
calculation to the $D$-dimensional black hole solution. Consider a
$D$-dimensional model of gravity coupled to a nonlinear electromagnetic field
$A_{\mu}$ with the action given by%
\begin{equation}
S_{\text{Bulk}}=\int d^{D}x\sqrt{-g}\left[  \frac{R-2\Lambda}{16\pi}%
+\frac{\mathcal{L}\left(  s;a_{i}\right)  }{4\pi}\right]  , \label{eq:Action}%
\end{equation}
where the cosmological constant $\Lambda=-\frac{\left(  D-1\right)  \left(
D-2\right)  }{2l^{2}}$, and $l$ is the AdS radius. In the action $\left(
\ref{eq:Action}\right)  $, we assume that the generic NLED Lagrangian
$\mathcal{L}\left(  s;a_{i}\right)  $ is a function of $s$ and the
dimensionful parameters $a_{i}$. The parameters $a_{i}$ characterize the
effects of nonlinearity in the NLED, and $s$ is an independent nontrivial
scalar using $F_{\mu\nu}=\partial_{\mu}A_{\nu}-\partial_{\nu}A_{\mu}$ and none
of its derivatives:
\begin{equation}
s=-\frac{1}{4}F^{\mu\nu}F_{\mu\nu}\text{.}%
\end{equation}
Varying the action $\left(  \ref{eq:Action}\right)  $ with respect to
$g_{\mu\nu}$ and $A_{\mu}$, we find that the equations of motion are%
\begin{align}
R_{\mu\nu}-\frac{1}{2}Rg_{\mu\nu}-\frac{\left(  D-1\right)  \left(
D-2\right)  }{2l^{2}}g_{\mu\nu}  &  =8\pi T_{\mu\nu}\text{,}\nonumber\\
\nabla_{\mu}G^{\mu\nu}  &  =0\text{,}%
\end{align}
where $T_{\mu\nu}$ is the energy-momentum tensor:%
\begin{equation}
T_{\mu\nu}=\frac{1}{4\pi}\left[  g_{\mu\nu}\mathcal{L}\left(  s;a_{i}\right)
+\frac{\partial\mathcal{L}\left(  s;a_{i}\right)  }{\partial s}F_{\mu}^{\text{
}\rho}F_{\nu\rho}\right]  \text{.}%
\end{equation}
Here, we define the auxiliary fields $G^{\mu\nu}$ for later convenience:%
\begin{equation}
G^{\mu\nu}\equiv-\frac{\partial\mathcal{L}\left(  s;a_{i}\right)  }{\partial
F_{\mu\nu}}=\frac{\partial\mathcal{L}\left(  s;a_{i}\right)  }{\partial
s}F^{\mu\nu}.
\end{equation}

To construct a charged AdS black hole solution, we take the following ansatz
for the metric and the NLED field%
\begin{align}
ds^{2}  &  =-f\left(  r\right)  dt^{2}+\frac{dr^{2}}{f\left(  r\right)
}+r^{2}d\Omega_{D-2}^{2}\text{,}\nonumber\\
A_{\mu}dx^{\mu}  &  =A_{t}\left(  r\right)  dt\text{,} \label{eq:ansatz}%
\end{align}
where $d\Omega_{D-2}^{2}$ is the metric on the unit $\left(  D-2\right)
$-sphere. The$\ $equations of motion for $tt$-component of gravity and
$A_{t}\left(  r\right)  $ are%
\begin{align}
-\frac{\left(  D-2\right)  \left(  D-3\right)  }{2}\frac{f\left(  r\right)
\left[  -1+f\left(  r\right)  -\frac{D-1}{D-3}\frac{r^{2}}{l^{2}}%
+\frac{rf^{\prime}\left(  r\right)  }{D-3}\right]  }{r^{2}}  &  =-2f\left(
r\right)  \left[  \mathcal{L}\left(  s;a_{i}\right)  +A_{t}^{\prime}%
G^{rt}\right]  \text{,}\label{eq:tt}\\
\left[  r^{D-2}G^{tr}\right]  ^{\prime}  &  =0\text{,} \label{eq:NLED}%
\end{align}
respectively, where%
\begin{equation}
s=\frac{A_{t}^{\prime2}\left(  r\right)  }{2}\text{ and }G^{tr}=\frac
{\partial\mathcal{L}\left(  s,a_{i}\right)  }{\partial s}A_{t}^{\prime}\left(
r\right)  \text{.} \label{eq:sGrt}%
\end{equation}
It can show that the rest equations of motion can be derived from eqns.
$\left(  \ref{eq:tt}\right)  $ and $\left(  \ref{eq:NLED}\right)  $. Solving
eqn. $\left(  \ref{eq:NLED}\right)  $ gives%
\begin{equation}
G^{tr}=\frac{\tilde{Q}}{r^{D-2}}\text{,} \label{eq:Grt}%
\end{equation}
where $\tilde{Q}$ is a constant. From eqns. $\left(  \ref{eq:sGrt}\right)  $
and $\left(  \ref{eq:Grt}\right)  $, $A_{t}^{\prime}\left(  r\right)  $ is
determined by%
\begin{equation}
\mathcal{L}^{\prime}\left(  \frac{A_{t}^{\prime2}\left(  r\right)  }{2}%
;a_{i}\right)  A_{t}^{\prime}\left(  r\right)  =\frac{\tilde{Q}}{r^{D-2}}.
\label{eq:QAt}%
\end{equation}
Furthermore, integrating eqn. $\left(  \ref{eq:tt}\right)  $ leads to
\begin{equation}
f\left(  r\right)  =1-\frac{\tilde{M}}{r^{D-3}}+\frac{r^{2}}{l^{2}}-\frac
{4}{D-2}\frac{1}{r^{D-3}}\int_{r}^{\infty}r^{D-2}\left[  \mathcal{L}\left(
s;a_{i}\right)  -A_{t}^{\prime}\left(  r\right)  \frac{\tilde{Q}}{r^{D-2}%
}\right]  dr, \label{eq:f(r)}%
\end{equation}
where $\tilde{M}$ is a constant. For large values of $r$, one finds that%
\begin{equation}
f\left(  r\right)  \approx1-\frac{\tilde{M}}{r^{D-3}}+\frac{r^{2}}{l^{2}%
}+\frac{2\tilde{Q}^{2}}{\left(  D-2\right)  \left(  D-3\right)  r^{2\left(
D-3\right)  }}, \label{eq:f(r)LV}%
\end{equation}
which reduces to the metric of a $D$-dimensional RN-AdS black hole. The
Hawking temperature of the black hole is given by%
\begin{equation}
T=\frac{f^{\prime}\left(  r_{+}\right)  }{4\pi}\text{.}%
\end{equation}
At $r=r_{+}$, eqn. $\left(  \ref{eq:tt}\right)  $ gives%
\begin{equation}
T=\frac{\left(  D-3\right)  }{4\pi r_{+}}\left\{  1+\frac{D-1}{D-3}\frac
{r_{+}^{2}}{l^{2}}+\frac{4r_{+}^{2}}{\left(  D-2\right)  \left(  D-3\right)
}\left[  \mathcal{L}\left(  \frac{A_{t}^{\prime2}\left(  r_{+}\right)  }%
{2};a_{i}\right)  -A_{t}^{\prime}\left(  r_{+}\right)  \frac{\tilde{Q}}%
{r_{+}^{D-2}}\right]  \right\}  . \label{eq:HT}%
\end{equation}

Since the black hole solution $\left(  \ref{eq:f(r)}\right)  $ is
asymptotically AdS, the asymptotic behavior of $f\left(  r\right)  $ in eqn.
$\left(  \ref{eq:f(r)LV}\right)  $ can relate the constant $\tilde{M}$ to the
black hole mass $M$ \cite{NLEDBH-Jamsin:2007qh},%
\begin{equation}
M=\frac{D-2}{16\pi}\omega_{D-2}\tilde{M}\text{,}%
\end{equation}
with $\omega_{D-2}$ being the volume of the unit $\left(  D-2\right)
$-sphere:%
\begin{equation}
\omega_{D-2}=\frac{2\pi^{\frac{D-1}{2}}}{\Gamma\left(  \frac{D-1}{2}\right)
}.
\end{equation}
The black hole charge $Q$ can be expressed in terms of the constants
$\tilde{Q}$. To do so, turning on an external current $J^{\mu}$ gives an
interaction term:
\begin{equation}
S_{I}=\int d^{4}x\sqrt{-g}J^{\mu}A_{\mu}.
\end{equation}
The equation of motion for $A_{\mu}$ then becomes%
\begin{equation}
\nabla_{\nu}G^{\mu\nu}=4\pi J^{\mu}. \label{eq:GmnJ}%
\end{equation}
The charge passing through a spacelike $t$-constant hypersurface $\Sigma_{t}$
is given by%
\begin{equation}
Q=-\int_{\Sigma_{t}}J^{\mu}n_{\mu}d\Sigma,,
\end{equation}
where $n_{\mu}$ is the unit normal vector of $\Sigma_{t}$, and $d\Sigma$ is
the surface element on $\Sigma_{t}$. Using Stokes' theorem, we rewrite $Q$ as
an integral over the boundary of $\Sigma_{t}$, i.e., the $\left(  D-2\right)
$-sphere $S^{D-2}$ at $r=\infty$,
\begin{equation}
Q=\frac{1}{8\pi}\oint\limits_{S^{D-2}}G^{\mu\nu}dS_{\mu\nu}=\frac{\tilde{Q}%
}{4\pi}\omega_{D-2}\text{,}%
\end{equation}
where eqns. $\left(  \ref{eq:Grt}\right)  $ and $\left(  \ref{eq:GmnJ}\right)
$ are used, and $dS_{\mu\nu}$ is the directed surface element on $S^{D-2}$.
The electrical potential measured with respect to the horizon is%
\begin{equation}
\Phi=\int_{r_{+}}^{\infty}A_{t}^{\prime}\left(  r\right)  =-A_{t}\left(
r_{+}\right)  , \label{eq:potential}%
\end{equation}
where we fix the field $A_{t}\left(  r\right)  $ at $r=\infty$ to be zero. The
electrostatic potential $\Phi$ plays a role as the conjugated variable to $Q$
in black hole thermodynamics. The entropy of the black hole is one-quarter of
the horizon area
\begin{equation}
S=\frac{r_{+}^{D-2}\omega_{D-2}}{4}\text{.} \label{eq:entropy}%
\end{equation}

Besides $s$, there exist other invariants built from $F_{\mu\nu}$ using the
totally antisymmetric Lorentz tensor $\epsilon^{\mu_{1}\cdots\mu_{D}}$. For
example, in four-dimensional spacetime, one can have another invariant
$p=-\frac{1}{8}\epsilon^{\mu\nu\rho\sigma}F_{\mu\nu}F_{\rho\sigma}$. The NLED
Lagrangian can be a function of $s$ and $p$. In fact, Born-Infeld
electrodynamics is described by the Lagrangian density%
\begin{equation}
\mathcal{L}_{\text{BI}}\left(  s,p;a\right)  =\frac{1}{a}\left(
1-\sqrt{1-2as-a^{2}p^{2}}\right)  \text{,} \label{eq:BI}%
\end{equation}
where the coupling parameter $a$ is related to the string tension
$\alpha^{\prime}$ as $a=\left(  2\pi\alpha^{\prime}\right)  ^{2}$. However,
when a black hole ansatz has no magnetic charge, the invariant $p$ always
vanishes, and hence $\mathcal{L}_{\text{BI}}\left(  s,p;a\right)  $ and
$\mathcal{L}_{\text{BI}}\left(  s,0;a\right)  $ would give the same solution.
Therefore, focusing on black hole solutions that are not magnetically charged,
our results also apply to the NLED with a more general Lagrangian containing
not only $s$ but also other invariants that vanish for zero magnetic charge.

\section{Thermodynamics via Charged Particle Absorption}

\label{Sec:TCPA}

In this section, we investigate thermodynamics of the NLED black hole solution
$\left(  \ref{eq:ansatz}\right)  $ by a charged particle entering the horizon.
When the particle is absorbed by the black hole, the mass and charge of the
black hole are varied due to the energy and charge conservation. Other
thermodynamic variables (e.g., the entropy) would change accordingly. The aim
of this section is to check whether the changes of the black hole
thermodynamic variables satisfy the first and second law of thermodynamics in
the normal and extended phase spaces. To do so, we first present some useful
formulae:%
\begin{align}
\frac{\partial f\left(  r\right)  }{\partial r}|_{r=r_{+}}  &  =4\pi T\text{,
}\nonumber\\
\frac{\partial f\left(  r\right)  }{\partial M}|_{r=r_{+}}  &  =-\frac{16\pi
}{\left(  D-2\right)  \omega_{D-2}r_{+}^{D-3}}\text{,}\nonumber\\
\frac{\partial f\left(  r\right)  }{\partial l}|_{r=r+}  &  =-\frac{2r_{+}%
^{2}}{l^{3}}\text{,}\label{eq:fdrh}\\
\frac{\partial f\left(  r\right)  }{\partial Q}|_{r=r_{+}}  &  =\frac
{16\pi\Phi}{\left(  D-2\right)  \omega_{D-2}r_{+}^{D-3}}\text{,}\nonumber\\
\frac{\partial f\left(  r\right)  }{\partial a_{i}}|_{r=r_{+}}  &
=\frac{16\pi\mathcal{A}_{i}}{\left(  D-2\right)  \omega_{D-2}r_{+}^{D-3}%
}\text{,}\nonumber
\end{align}
where $\mathcal{A}_{i}$ is defined in eqn. $\left(  \ref{eq:A}\right)  $, and
eqn. $\left(  \ref{eq:QAt}\right)  $ are used to obtain $\partial f\left(
r\right)  /\partial Q|_{r=r_{+}}$ and $\partial f\left(  r\right)  /\partial
a|_{r=r_{+}}$.

\subsection{Normal Phase Space}

In the normal phase space, the AdS radius $l$ and the parameters $a_{i}$ are
fixed and not treated as thermodynamic variables. So the black hole mass $M$
is the internal energy of the black hole. After the black hole absorbs a
charged particle of the energy $E$ and charge $q$, the state of the black hole
changes from $\left(  M,Q\right)  $ to $\left(  M+dM,Q+dQ\right)  $. The
energy and charge conservation of the absorbing process gives%
\begin{equation}
dM=E\text{ and }dQ=q\text{,} \label{eq:dMdQNS}%
\end{equation}
where $E$ and $q$ are related via eqn. $\left(  \ref{eq:Ehorizon}\right)  $.
To treat the particle as a test particle, we shall assume its energy $E$ and
charge $q$ are small compared to those of the black hole,%
\begin{equation}
q=dQ\ll Q\text{ and }E=dM\text{ }\ll M\text{.}%
\end{equation}
For the initial state $\left(  M,Q\right)  $, the outer horizon radius of the
black hole $r_{+}$ satisfies%
\begin{equation}
f\left(  r_{+};M,Q\right)  =0\text{,}%
\end{equation}
where the parameters $M$ and $Q$ are temporarily put explicitly as arguments
of the function $f\left(  r\right)  $. When the black hole mass and charge are
varied, the outer horizon radius moves to $r_{+}+dr_{+}$, which also satisfies%
\begin{equation}
f\left(  r_{+}+dr_{+};M+dM,Q+dQ\right)  =0\text{.}%
\end{equation}
So the infinitesimal changes in $M$, $Q$ and $r_{+}$ are related by%
\begin{equation}
\frac{\partial f\left(  r\right)  }{\partial r}|_{r=r_{+}}dr_{+}%
+\frac{\partial f\left(  r\right)  }{\partial M}|_{r=r_{+}}dM+\frac{\partial
f\left(  r\right)  }{\partial Q}|_{r=r_{+}}dQ=0\text{.} \label{eq:NPFL}%
\end{equation}
Using eqns. $\left(  \ref{eq:Ehorizon}\right)  $, $\left(  \ref{eq:entropy}%
\right)  $ and $\left(  \ref{eq:fdrh}\right)  $, we find that the above
equation reduces to
\begin{equation}
\left\vert P^{r}\left(  r_{+}\right)  \right\vert =TdS\text{.}
\label{eq:TDSNS}%
\end{equation}
Therefore, the variation of entropy is%
\begin{equation}
dS=\frac{\left\vert P^{r}\left(  r_{+}\right)  \right\vert }{T}>0\text{,}%
\end{equation}
which means the second law of thermodynamics is always satisfied. Substituting
eqn. $\left(  \ref{eq:TDSNS}\right)  $ into eqn. $\left(  \ref{eq:Ehorizon}%
\right)  $, we obtain%
\begin{equation}
dM=\Phi dQ+TdS\text{,} \label{eq:1stNS}%
\end{equation}
which is the first law of thermodynamics.

\subsection{Extended Phase Space}

Comparing with the first law of a usual thermodynamic system, one notes that
there is an omission of a pressure volume term $PdV$ in eqn. $\left(
\ref{eq:1stNS}\right)  $. This observation motivates treating the cosmological
constant as pressure associated with a black hole
\cite{IN-Kastor:2009wy,IN-Dolan:2011xt}:%
\begin{equation}
P\equiv-\frac{\Lambda}{8\pi}=\frac{\left(  D-1\right)  \left(  D-2\right)
}{16\pi l^{2}}.
\end{equation}
In this case, the mass of the black hole is no longer identified with internal
energy, rather it is regarded as a gravitational version of chemical enthalpy.
So the conjugate thermodynamic volume of the black hole is given by%
\begin{equation}
V=\frac{\partial M}{\partial P}\text{.}%
\end{equation}
Using eqn. $\left(  \ref{eq:f(r)}\right)  $ for $f\left(  r\right)  $ and
$f\left(  r_{+}\right)  =0$, we find that the thermodynamic volume of the NLED
black hole is
\begin{equation}
V=\frac{\omega_{D-2}}{D-1}r_{+}^{D-1}\text{.}%
\end{equation}

To satisfy the Smarr relation, we need to further enlarge the phase space by
promoting any dimensionful parameter in the theory to a thermodynamic variable
\cite{IN-Gunasekaran:2012dq,TCPA-Kastor:2010gq}, which introduce the
associated conjugate. The parameters and their conjugates add extra terms in
the first law of thermodynamics. In particular, the parameter $a_{i}$ in the
NLED theory $\left(  \ref{eq:Action}\right)  $ is considered as a
thermodynamic phase space variable, and the associated conjugates
$\mathcal{A}_{i}$ is%
\begin{equation}
\mathcal{A}_{i}=\frac{\partial M}{\partial a_{i}}=-\frac{\omega_{D-2}}{4\pi
}\int_{r_{+}}^{\infty}r^{D-2}\frac{\partial\mathcal{L}\left(  s;a_{i}\right)
}{\partial a_{i}}dr\text{.} \label{eq:A}%
\end{equation}
In \cite{IN-Wang:2018xdz}, we showed that a complete Smarr relation is
satisfied by all thermodynamic quantities in the extended phase space,%
\begin{equation}
M=2\left(  TS-VP\right)  +\sum\nolimits_{i}c_{i}a_{i}\mathcal{A}_{i}%
+Q\Phi\text{,}%
\end{equation}
where $\left[  a_{i}\right]  =L^{c_{i}}$.

After a charged particle of the energy $E$ and charge $q$ enters the black
hole horizon, the internal energy and charge of the black hole are changed by%
\begin{equation}
d\left(  M-PV\right)  =E\text{ and }dQ=q\text{,} \label{eq:EQES}%
\end{equation}
where $M-PV$ is the black hole internal energy in the extended phase space.
Thus, the black hole changes from the initial state $\left(  M,Q,l,a_{i}%
\right)  $ to the final state $\left(  M+dM,Q+dQ,l+dl,a_{i}+da_{i}\right)  $.
The infinitesimal changes in $M$, $Q$, $l$, $a_{i}$ and $r_{+}$ are related by%
\begin{equation}
\frac{\partial f\left(  r\right)  }{\partial r}|_{r=r_{+}}dr_{+}%
+\frac{\partial f\left(  r\right)  }{\partial M}|_{r=r_{+}}dM+\frac{\partial
f\left(  r\right)  }{\partial Q}|_{r=r_{+}}dQ+\frac{\partial f\left(
r\right)  }{\partial l}|_{r=r_{+}}dl+\frac{\partial f\left(  r\right)
}{\partial a_{i}}|_{r=r_{+}}da_{i}=0\text{.} \label{eq:deltaES}%
\end{equation}
From eqn. $\left(  \ref{eq:deltaES}\right)  $, we find that%
\begin{equation}
\left\vert P^{r}\left(  r_{+}\right)  \right\vert =TdS+\sum\nolimits_{i}%
\mathcal{A}_{i}da_{i}-PdV, \label{eq:PrES}%
\end{equation}
where we use eqn. $\left(  \ref{eq:Ehorizon}\right)  $ to express $dQ$ in
terms of $\left\vert P^{r}\left(  r_{+}\right)  \right\vert $, and the
derivatives of $f\left(  r\right)  $ are given in $\left(  \ref{eq:fdrh}%
\right)  $. Plugging the above equation into eqn. $\left(  \ref{eq:Ehorizon}%
\right)  $, we use eqn. $\left(  \ref{eq:EQES}\right)  $ to obtain%
\begin{equation}
dM=\Phi dQ+TdS+VdP+\sum\nolimits_{i}\mathcal{A}_{i}da_{i}\text{,}%
\end{equation}
which is the first law of thermodynamics.

Eqn. $\left(  \ref{eq:PrES}\right)  $ gives the change of the black hole entropy,%

\begin{equation}
dS=\frac{\left\vert P^{r}\left(  r_{+}\right)  \right\vert -\sum
\nolimits_{i}\mathcal{A}_{i}da_{i}}{T-\frac{\left(  D-1\right)  r_{+}}{4\pi
l^{2}}}\text{.} \label{eq:dsES}%
\end{equation}
For large values of $T$, the denominator in eqn. $\left(  \ref{eq:dsES}%
\right)  $ becomes%
\begin{equation}
T-\frac{\left(  D-1\right)  r_{+}}{4\pi l^{2}}\sim\frac{\left(  D-3\right)
}{4\pi r_{+}}>0,
\end{equation}
where we use eqn. $\left(  \ref{eq:HT}\right)  $. However for small enough
$T$, the denominator is negative. Since $da_{i}$ is arbitrary, the sign of the
numerator in eqn. $\left(  \ref{eq:dsES}\right)  $ is indefinite. In the
extended phase space, the entropy can increases or decrease depending the
values of $da_{i}$. In the \textquotedblleft restricted\textquotedblright%
\ extended phase space with $da_{i}=0$, the change of the black hole entropy
becomes%
\begin{equation}
dS=\frac{\left\vert P^{r}\left(  r_{+}\right)  \right\vert }{T-\frac{\left(
D-1\right)  r_{+}}{4\pi l^{2}}}\text{,}%
\end{equation}
which shows that the second law of thermodynamics is not satisfied for the
extremal or near-extremal black hole. When the black hole is far enough from
extremality, the second law is satisfied in the \textquotedblleft
restricted\textquotedblright\ extended phase space.

\section{Weak Cosmic Censorship Conjecture}

\label{Sec:WCCC}

To test the weak cosmic censorship conjecture, we make the black hole, which
is initially extremal or near extremal, absorb a charged particle and check
whether the resulting object has too much charge to be a black hole. Since the
particle is a test particle, the black hole needs to start out close to
extremal to have chance to become a naked singularity. So here we assume that
the initial NLED black hole is near extremal, for which there are two
horizons. Between these two horizons, there exists one and only one minimum
point at $r=r_{\min}$ for $f\left(  r\right)  $. Moreover, the minimum value
of $f\left(  r\right)  $ is not greater than zero,%
\begin{equation}
\delta\equiv f\left(  r_{\min}\right)  \leq0\text{,}%
\end{equation}
and $\delta=0$ corresponds to the extremal black hole. After the black hole
absorbs a charged particle, the minimum point would move to $r_{\min}%
+dr_{\min}$. For the final black hole solution, if the minimum value of
$f\left(  r\right)  $ at $r=r_{\min}+dr_{\min}$ is still not greater than
zero, there is an event horizon. Otherwise, the final black hole solution is
over the extremal limit, and the weak cosmic censorship conjecture is
violated. Again, we first present some useful formulae:%
\begin{align}
\frac{\partial f\left(  r\right)  }{\partial r}|_{r=r_{\min}}  &  =0\text{,
}\nonumber\\
\frac{\partial f\left(  r\right)  }{\partial M}|_{r=r_{\min}}  &
=-\frac{16\pi}{\left(  D-2\right)  \omega_{D-2}r_{\min}^{D-3}}\text{,}%
\nonumber\\
\frac{\partial f\left(  r\right)  }{\partial l}|_{r=r_{\min}}  &
=-\frac{2r_{\min}^{2}}{l^{3}}\text{,}\label{eq:fdrmin}\\
\frac{\partial f\left(  r\right)  }{\partial Q}|_{r=r_{\min}}  &  =\frac
{16\pi\left[  \Phi+A_{t}\left(  r_{+}\right)  -A_{t}\left(  r_{\min}\right)
\right]  }{\left(  D-2\right)  \omega_{D-2}r_{\min}^{D-3}}\text{,}\nonumber\\
\frac{\partial f\left(  r\right)  }{\partial a_{i}}|_{r=r_{\min}}  &
=\frac{16\pi\left(  \mathcal{A}_{i}+\delta\mathcal{A}_{i}\right)  }{\left(
D-2\right)  \omega_{D-2}r_{\min}^{D-3}}\text{,}\nonumber
\end{align}
where eqn. $\left(  \ref{eq:QAt}\right)  $ are used to obtain $\partial
f\left(  r\right)  /\partial Q|_{r=r_{\min}}$ and $\partial f\left(  r\right)
/\partial a_{i}|_{r=r_{\min}}$, and we define%
\begin{equation}
\delta\mathcal{A}_{i}\equiv-\frac{\omega_{D-2}}{4\pi}\int_{r_{\min}}^{r_{+}%
}drr^{2}\frac{\partial\mathcal{L}\left(  s;a_{i}\right)  }{\partial a_{i}%
}\text{.}%
\end{equation}

\subsection{Normal Phase Space}

In this case, absorbing a charged particle of the energy $E$ and charge $q$
makes the black hole change from the initial state $\left(  M,Q\right)  $ to
the final state $\left(  M+dM,Q+dQ\right)  $, where $dM$ and $dQ$ are given in
eqn. $\left(  \ref{eq:dMdQNS}\right)  $. Owing to these changes, the location
of the minimum point of $f\left(  r\right)  $ moves to $r_{\min}+dr_{\min}$.
For the final state, the minimum value of $f\left(  r\right)  $ at $r=r_{\min
}+dr_{\min}$ becomes%
\begin{align}
&  f\left(  r_{\min}+dr_{\min};M+dM,Q+dQ\right) \nonumber\\
&  =\delta+\frac{\partial f}{\partial r}|_{r=r_{\min}}dr_{\min}+\frac{\partial
f}{\partial M}|_{r=r_{\min}}dM+\frac{\partial f}{\partial Q}|_{r=r_{\min}%
}dQ\label{eq:fminNS}\\
&  =\delta-\frac{16\pi\left\vert P^{r}\left(  r_{+}\right)  \right\vert
}{\left(  D-2\right)  \omega_{D-2}r_{\min}^{D-3}}+\frac{16\pi q\left[
A_{t}\left(  r_{+}\right)  -A_{t}\left(  r_{\min}\right)  \right]  }{\left(
D-2\right)  \omega_{D-2}r_{\min}^{D-3}},\nonumber
\end{align}
where the parameters of $f\left(  r\right)  $ are put explicitly. If the
initial black hole is extremal, we have $r_{+}=r_{\min}$ and $\delta=0$. The
minimum value of $f\left(  r\right)  $ of the final black hole reduces to%
\begin{equation}
f\left(  r_{\min}+dr_{\min}\right)  =-\frac{16\pi\left\vert P^{r}\left(
r_{+}\right)  \right\vert }{\left(  D-2\right)  \omega_{D-2}r_{\min}^{D-3}%
}<0\text{,}%
\end{equation}
which implies that the extremal black hole becomes a non-extremal one by the
absorption. For the near-extremal black hole, we define $\epsilon$ such that%
\begin{equation}
r_{\min}=r_{+}\left(  1-\epsilon\right)  \text{,}%
\end{equation}
where $\epsilon\ll1$. So $\delta$ is suppressed by $\epsilon$ in the
near-extremal limit. Moreover, the second term in the third line of eqn.
$\left(  \ref{eq:fminNS}\right)  $ is only suppressed by the test particle
limit while the third term is suppressed by both near-extremal limit and the
test particle limit. Therefore, in the test particle limit for the
near-extremal black hole, the third term can be neglected, and eqn. $\left(
\ref{eq:fminNS}\right)  $ then gives%
\begin{equation}
f\left(  r_{\min}+dr_{\min}\right)  =\delta-\frac{16\pi\left\vert P^{r}\left(
r_{+}\right)  \right\vert }{\left(  D-2\right)  \omega_{D-2}r_{\min}^{D-3}%
}<0\text{,}%
\end{equation}
which means that the near-extremal black hole stays non-extremal after the
absorption. In the normal phase space, the weak cosmic censorship conjecture
is satisfied for the extremal and near-extremal NLED black holes under the
charge particle absorption.

\subsection{Extended Phase Space}

In the extended phase space, after absorbing a charged particle, the
parameters of the black hole change from $\left(  M,Q,l,a_{i}\right)  $ to
$\left(  M+dM,Q+dQ,l+dl,a_{i}+da_{i}\right)  $, and $r_{\min}$ moves
accordingly to $r_{\min}+dr_{\min}$. The minimum value of $f\left(  r\right)
$ at $r=r_{\min}+dr_{\min}$ of the final state is given by%
\begin{align}
&  f\left(  r_{\min}+dr_{\min};M+dM,Q+dQ,l+dl,a_{i}+da_{i}\right) \nonumber\\
&  =\delta+\frac{\partial f}{\partial M}|_{r=r_{\min}}dM+\frac{\partial
f}{\partial Q}|_{r=r_{\min}}dQ+\frac{\partial f}{\partial l}|_{r=r_{\min}%
}dl+\sum\nolimits_{i}\frac{\partial f}{\partial a_{i}}|_{r=r_{\min}}%
da_{i}\label{eq:fminES}\\
&  =\delta-\frac{16\pi\left\{  TdS+\left[  A_{t}\left(  r_{+}\right)
-A_{t}\left(  r_{\min}\right)  \right]  q+\delta\mathcal{A}_{i}da_{i}\right\}
}{\left(  D-2\right)  \omega_{D-2}r_{\min}^{D-3}}+2\left(  \frac{r_{+}^{D-1}%
}{r_{\min}^{D-3}}-r_{\min}^{2}\right)  \frac{dl}{l^{3}},\nonumber
\end{align}
where the parameters of $f\left(  r\right)  $ are put explicitly, and we use
eqn. $\left(  \ref{eq:fdrmin}\right)  $ for the derivatives of $f\left(
r\right)  $. When the initial black hole is extremal, one has that
$r_{+}=r_{\min}$, $T=0$, $\delta\mathcal{A}_{i}=0$ and $\delta=0$. So the
minimum value of $f\left(  r\right)  $ of the final black hole is%
\begin{equation}
f\left(  r_{\min}+dr_{\min}\right)  =0\text{,}%
\end{equation}
which implies that the extremal black hole stays extremal after the
absorption. For the near-extremal black hole in the test particle limit, the
second and third terms in the third line of eqn. $\left(  \ref{eq:fminES}%
\right)  $ are suppressed by both the near-extremal limit and the test
particle limit and hence can be neglected. Therefore, eqn. $\left(
\ref{eq:fminES}\right)  $ leads to%
\begin{equation}
f\left(  r_{\min}+dr_{\min}\right)  =\delta<0\text{,}%
\end{equation}
which means that the near-extremal black hole stays non-extremal after the
absorption. In the extended phase space, the weak cosmic censorship conjecture
is also satisfied for the extremal and near-extremal NLED black holes under
the charge particle absorption.

\section{Conclusion}

\label{Sec:Con}

In this paper, we first derived the $D$-dimensional asymptotically AdS charged
black hole solution for the general NLED. Then, we calculated the variations
of the thermodynamic quantities of a general NLED-AdS black hole via absorbing
a charged practice. With these variations, we checked the first and second
laws of thermodynamics and the weak cosmic censorship conjecture for the NLED
black hole in the test particle limit. Two scenarios were considered, namely
the normal phase space and the extended phase space. Our results are
summarized in Table \ref{tab:1}.

\begin{table}[tbh]
\centering%
\begin{tabular}
[c]{|p{0.6in}|p{2.6in}|p{2.6in}|}\hline
& Norma Phase Space & Extended Phase Space\\\hline
1st Law & Satisfied. & Satisfied.\\\hline
2nd Law & Satisfied. & Indefinite. If $da_{i}=0$ is imposed, the 2nd law is
violated for the extremal and near-extremal black holes.\\\hline
WCCC & Satisfied for the extremal and near-extremal black holes. After the
charge particle absorption, the extremal black hole becomes non-extremal. &
Satisfied for the extremal and near-extremal black holes. After the charge
particle absorption, the extremal black hole stays extremal.\\\hline
\end{tabular}
$\ \ $\caption{{\small Results for the first and second laws of thermodynamics
and the weak cosmic censorship conjectures (WCCC), which are tested for a
general NLED charged AdS black hole with absorbing a charged particle in the
test particle limit.}}%
\label{tab:1}%
\end{table}

In \cite{IN-Gwak:2017kkt}, the first and second laws of thermodynamics and the
weak cosmic censorship conjecture have been tested for the Einstein-Maxwell
action in the $D$-dimensional spacetime, which is the special case of the
Einstein-NLED theory, i.e., $\mathcal{L}\left(  s;a_{i}\right)  =s$. The
analysis was considered in the extended phase space, in which there is no
dimensionful parameter in the Maxwell Lagrangian. As expected, our results
successfully reproduced those in \cite{IN-Gwak:2017kkt}. Moreover, our results
show that the correction terms to the electromagnetic part can not solve the
violation of the second law of thermodynamics near extremality, and hence the
correction terms to the gravity part should be considered. In
\cite{IN-Zeng:2019jta}, the thermodynamics and the weak cosmic censorship
conjecture of the Born-Infeld AdS black holes were discussed in the normal and
extended phase space. Our results agree with those in the normal phase space.
However, in the extended space, it showed there that the extremal black holes
will change into non-extremal black holes after the charge particle
absorption, which can not reduce to the Maxwell case in \cite{IN-Gwak:2017kkt}
with $b=0$. Recently, a $\left(  2+1\right)  $-dimensional regular black hole
with some nonlinear electrodynamic source was considered in
\cite{CON-Han:2019lfs}, the results of which agree with ours.

\begin{acknowledgments}
We are grateful to Zheng Sun for useful discussions. This work is supported in
part by NSFC (Grant No. 11005016, 11875196 and 11375121).
\end{acknowledgments}

\end{document}